\documentclass[reprint,preprintnumbers,amsmath,amssymb,nofootinbib,aps,prx]{revtex4-2}

\usepackage{slashed}
\usepackage{amsmath}
\usepackage{amssymb}
\usepackage{amsthm}
\usepackage{hyperref}
\usepackage{indentfirst}
\usepackage{psfrag}
\usepackage{graphicx}
\usepackage[left=20mm,right=20mm,top=20mm,bottom=20mm]{geometry}
\usepackage{cleveref}
\usepackage{xcolor}

\hypersetup{colorlinks=true, linkcolor=blue, urlcolor=blue, citecolor=blue, linktocpage=true}

\usepackage{xifthen}% Provides \isempty test
\newcommand*\diff{\mathrm{d}} % Straight differential
\newcommand*\ldiff[2][]{ \ifthenelse{\isempty{#1}}{ \diff #2}{\diff^#1#2} \,} % Differential with measure; the mandatory argument is the name of the measure, the option one is the dimension
\let\limitint\int % Only when I provide explicit limits for the integration, I need to do the spacing myself
\renewcommand{\int}{\limitint \!} % The standard integral should have correct spacing

\def\e{{\rm e}}

\newcommand{\bseq}{\begin{subequations}}
\newcommand{\eseq}{\end{subequations}}

\newcommand{\Tr}{{\rm Tr}}

\newcommand{\be}{\begin{equation}}
\newcommand{\ee}{\end{equation}}
\newcommand{\beqa}{\begin{eqnarray}}
\newcommand{\eeqa}{\end{eqnarray}}

\renewcommand{\L}{\mathcal{L}}

\begin{document}

\preprint{INR-TH-2024-005}

\title{ Nontopological Electromagnetic Hedgehogs }

\author{Yulia Galushkina$^1$}
\email{jgrabareva@gmail.com}

\author{Emin Nugaev$^1$}
\email{emin@ms2.inr.ac.ru}

\author{Andrey Shkerin$^2$}
\email{ashkerin@perimeterinstitute.ca}

\affiliation{$^1$ Institute for Nuclear Research of RAS, prospekt 60-letiya Oktyabrya 7a, Moscow, 117312, Russia}
\affiliation{$^2$ Perimeter Institute for Theoretical Physics, Waterloo, Ontario, N2L 2Y5, Canada}

\begin{abstract}

We study classical localised configurations---solitons---in a theory of self-interacting complex Proca field with the global $U(1)$ symmetry.
We focus on spherically-symmetric solitons near the nonrelativistic limit, which are supported by the quartic interactions of the neutral Proca field.
Such solitons can source the radial electric (magnetic) field if one introduces a parity-even (parity-odd) coupling of the Proca field to the electromagnetic field tensor.
We discuss the conditions of existence of such nontopological ``electromagnetic hedgehogs'' and their properties.

\end{abstract}

\maketitle

\section{Introduction}

Nontopological solitons \cite{Lee:1991ax} are stationary nonlinear particle-like solutions of field equations in theories with global or local symmetries.
They exist due to self-interaction of the field, gravitational attraction, or some background potential. 
The best-known example of nontopological solitons is Q-balls \cite{Rosen:1968mfz,Coleman:1985ki} arising in theories of complex scalar field with the global or local \cite{Rosen:1968zwl,Lee:1988ag} $U(1)$ symmetry.
The symmetry leads to the conserved charge that can prevent the Q-ball, whose stability is not guaranteed by topology, from dissolving into free particles.
Q-balls found numerous applications in particle physics and cosmology and provide valuable insights into behaviour of other solutions of classical equations of motion \cite{Nugaev:2019vru}.

Analogs of Q-balls exist in theories of self-interacting complex massive vector (Proca) field \cite{Loginov:2015rya}.
The vector Q-balls, self-gravitating Proca stars \cite{Brito:2015pxa}---analogs of scalar Boson stars \cite{Liebling:2012fv,Visinelli:2021uve}, and their gauged counterparts \cite{SalazarLandea:2016bys} have been extensively studied in recent years \cite{Brihaye:2017inn,Minamitsuji:2018kof,Heeck:2021bce,Aoki:2022mdn,Wang:2023tly}.
The motivation comes partially from cosmology, where massive vector particles can be a dark matter candidate \cite{Antypas:2022asj}, and spatially localised structures made of these particles are exotic compact objects \cite{Cardoso:2019rvt}.

Theories of self-interacting vector field are effective field theories.
In the exceptional case of non-Abelian gauge theory, the well-known weakly coupled UV completion involves the Higgs mechanism, but this approach is not applicable in the general case.\footnote{For example, spin-1 atoms and $\rho$-mesons are composite particles in the complete strongly coupled theories.} Moreover, as was shown in \cite{Mou:2022hqb,Coates:2022qia,Coates:2023swo}, pathologies in models of self-interacting Proca field may appear already at the classical level. This does not preclude viewing relativistic massive vector theories as low-energy effective theories. Solitons and other classical configurations can be studied in this framework \cite{Lee:1994sk,Herdeiro:2023lze}, which we adopt in this paper.

%{\bf Although spin-1 particles and the corresponding vector fields are observable (e.g, spin-1 atoms, W-bosons, $\rho$-mesons), there are several obstacles for the construction of UV-complete quantum theory of interacting vector fields. For the exceptional case of non-Abelian gauge theory, there is a well-known way to complete the theory of massive vector bosons using the Higgs mechanism, but in the general case this approach is not applicable. Moreover, as was shown in \cite{Mou:2022hqb,Coates:2022qia,Coates:2023swo}, the pathologies in the models with self-interacting Proca fields may appear even at the classical level. Due to all these issues, the massive vector theories are usually considered as low-energy effective theories\footnote{For example, atoms and $\rho$-mesons are the composite particles in the completed strongly-coupled theories.}. In this framework, one can consider solitons \cite{Lee:1994sk,Herdeiro:2023lze} and other classical configurations. 
%}

%Theories of self-interacting vector field are, generally, effective field theories; the dynamics of the field \cite{Mou:2022hqb,Coates:2022qia,Coates:2023swo} and of corresponding solitons may be sensitive to the scale at which the effective theory breaks down and a UV completion is needed \cite{Aoki:2022woy,Herdeiro:2023lze}.

In this paper we study non-gravitating solitons in the theory of massive complex vector field $V^\mu$ with quartic self-interactions.
The theory possesses the global $U(1)$ symmetry which ensures the existence of solitons.
Interestingly, one can introduce a coupling of $V^\mu$ to the electromagnetic field tensor $F_{\mu\nu}$.
Such coupling may arise as the low-energy limit of the gauge- and global $U(1)$-invariant coupling between $V^\mu$ and the gauge field of the $U(1)_Y$-group of the Standard Model.
Depending on the parity of the coupling term, the interaction will endow $V^\mu$-particles with the electric or magnetic dipole moment, and the vector soliton will act as a source of electric or magnetic field confined to the bulk of the soliton.
We will be interested in spherically-symmetric configurations giving rise to radially polarised fields.
These objects are somewhat similar to nontopological magnetic monopoles studied in \cite{Lee:1994sk}, although in our case the fields decay exponentially fast beyond the bulk of the soliton.

The paper is organised as follows.
In Sec.~\ref{sec:setup} we introduce the model and derive the field equations and conserved currents.
In Sec.~\ref{sec:sol} we introduce the field ansatz for solitons and discuss their general properties.
We then find the solitons numerically, paying special attention to the nonrelativistic limit and the regime where the thin-wall approximation is applicable.
We also touch upon the validity of the effective field theory description for our solutions.
In Sec.~\ref{sec:hedgehog} we add the electromagnetic field to the picture of the solitons, and in Sec.~\ref{sec:concl} we conclude.
Throughout the paper we put $\hbar=c=1$ and use the metric signature $(+,-,-,-)$.

\section{Setup}
\label{sec:setup}

Consider a (3+1)-dimensional theory of the complex massive neutral field $V^\mu$ coupled to the electromagnetic field tensor $F_{\mu\nu}$.
We require the global $U(1)$ symmetry of this theory and the gauge-invariant coupling of $V^\mu$ to $F_{\mu\nu}$.
The general parity (P)-even Lagrangian reads as
\begin{multline}
\label{L}
\mathcal{L} = -\frac{1}{4} F_{\mu\nu}F^{\mu\nu} -\frac{1}{2} V^*_{\mu\nu}V^{\mu\nu} + \\ + \frac{ i \gamma}{2}F_{\mu\nu}W^{\mu\nu} - U(V^\nu, V^{*\mu}) \;, 
\end{multline}
where $F_{\mu\nu} = \partial_\mu A_\nu - \partial_\nu A_\mu$, $V_{\mu\nu} = \partial_\mu V_\nu - \partial_\nu V_\mu$, and $W_{\mu\nu} =  V^*_\mu V_\nu - V^*_\nu V_\mu$. 
Furthermore, $\gamma$ is a dimensionless constant, and \(U(V^\nu, V^{*\mu})\) is a $U(1)$-invariant potential of self-interaction. 

The real interaction term $\frac{i\gamma}{2}F_{\mu\nu}W^{\mu\nu}$ was first introduced in \cite{Corben:1940zz}. In particle physics, this interaction describes three-vector-boson coupling which was considered in Electroweak phenomenology \cite{Hagiwara:1986vm} and studied in a low-energy effective theory of strongly interacting spin-1 particles \cite{Davoudi:2015zda}.
In the non-relativistic limit the term $\frac{i\gamma}{2}F_{\mu\nu}W^{\mu\nu}$ generates the magnetic dipole moment \((\gamma/2M)\) \cite{Pauli:1941zz,Lee:1962vm} and the electric quadrupole moment \( (- \gamma/M^2)\) \cite{Hagiwara:1986vm,Davoudi:2015zda} of a vector particle.

%{\bf In the non-relativistic limit the term $\frac{i\gamma}{2}F_{\mu\nu}W^{\mu\nu}$ generates the magnetic dipole moment \((\gamma/2M)\) \cite{Pauli:1941zz,Lee:1962vm} and the electric quadrupole moment \( (- \gamma/M^2)\) \cite{Hagiwara:1986vm,Davoudi:2015zda} of a vector particle.}
%The interaction of massive neutral vector bosons with the gauge field has %a clear interpretation in the non-relativistic limit. 
%Namely, the term $\frac{i\gamma}{2}F_{\mu\nu}W^{\mu\nu}$ generates the %magnetic dipole moment of a vector particle \cite{Pauli:1941zz,Lee:1962vm}. 
As a physical example, the classical field $V_\mu$ describes a gas of spin-1 atoms in a Bose-Einstein condensate \cite{Dalfovo:1999zz}.
In this case, the global $U(1)$ symmetry of the theory (\ref{L}) corresponds to the $U(1)_B$ symmetry for the baryon number conservation.
This motivates us to pay special attention to the solitons in the non-relativistic (low-energy) regime of the theory.
At the quantum level, the model (\ref{L}) is an effective theory that is valid below the cutoff scale $\sim M/\sqrt{\gamma}$ or a scale associated with the vector self-interaction. 
Note that the model can not be considered as a low-energy limit of some renormalizable non-Abelian gauge theory. One possibility for a UV completion is to match the model to some strongly coupled theory at low energies \cite{Davoudi:2015zda}.
We will follow the approach of \cite{Lee:1994sk,Pombo:2023sih, Herdeiro:2023lze} and study classical solutions in the model assuming that the coupling constants in (\ref{L}) are independent. Note also that if the massive vector field is a dark matter candidate, there is a strong bound on $\gamma$  \cite{Chu:2023zbo}, and the cutoff of the theory  is determined by the vector field self-interaction \cite{Aoki:2022mdn}.

%Note also that if the massive vector field is a dark matter candidate, there is a strong bound on $\gamma$ \cite{Barger:2010gv}, and the cutoff of the theory  is determined by the vector field self-interaction \cite{Aoki:2022mdn}.  

%the matching of strong-coupling theory  to the low-energy effective theory\cite{Davoudi:2015zda}. Following \cite{Lee:1994sk,Pombo:2023sih, Herdeiro:2023lze}, we will study solitons in the model (\ref{L}) using classical analysis and our dimensionless couplings are independent. 
%; see, e.g., \cite{Pombo:2023sih, Herdeiro:2023lze}. 

The Lagrangian (\ref{L}) leads to the Maxwell equations
\begin{equation}
\label{E_to_W}
\partial_\mu F^{\mu\nu} = i \gamma \partial_\mu W^{\mu\nu} \;,
\end{equation}
which, for configurations with trivial boundary conditions at spatial infinity, can be integrated to yield
\begin{equation}
\label{E_to_W2}
F^{\mu\nu} = i \gamma W^{\mu\nu} \;.
\end{equation}
Next, we choose the most general P-even 4th-order potential of self-interaction:
\begin{equation}
\label{U1}
U = - M^2V^*_{\mu}V^{\mu} -\frac{\alpha}{2}(V^*_{\mu}V^{\mu})^2 - \frac{\beta}{2}(V^*_{\mu}V^{*\mu})(V_{\nu}V^{\nu})
\end{equation}
with \(\alpha\), \(\beta\) dimensionless constants. 
Substituting \cref{E_to_W,U1} to (\ref{L}), we arrive at the Lagrangian containing the vector field only,
\begin{equation}
\begin{aligned}
\label{eff_model}
& \L =  -\frac{1}{2} V^*_{\mu\nu}V^{\mu\nu} - \tilde{U}(V^\nu, V^{*\mu}) \;, \\
& \tilde{U} = -  M^2V^*_{\mu}V^{\mu}  - \frac{\tilde{\alpha}}{2} (V^*_{\mu}V^{\mu})^2 - \frac{\tilde{\beta}}{2}(V^*_{\mu}V^{*\mu})(V_{\nu}V^{\nu}) \;, \\  
\end{aligned}
\end{equation}
where  \( \tilde{\alpha} = \alpha + \gamma^2\) and \( \tilde{\beta} = \beta - \gamma^2\).

Let us discuss the possibility of enhancing the Lagrangian (\ref{L}) with the P-odd term \(\frac{ i \Tilde{\gamma} }{2}\Tilde{F}_{\mu\nu}W^{\mu\nu}\), where \(\Tilde{F}_{\mu\nu} =  \varepsilon_{\mu \nu \rho \lambda} F^{\rho\lambda}  \) and \(\Tilde \gamma\) is a dimensionless coupling constant.
In the non-relativistic limit, this interaction provides the vector boson with the electric dipole and magnetic quadrupole moment.
Considering this coupling instead of \(\frac{ i \gamma }{2}F_{\mu\nu}W^{\mu\nu}\) leads again to the effective theory (\ref{eff_model}) with \( \tilde{\alpha} = \alpha -  4\Tilde{\gamma}^2\) and \( \tilde{\beta} = \beta + 4\Tilde{\gamma}^2\).
Note that both P-even and P-odd couplings provide the P-even contributions to the effective Lagrangian.
This is because $W_{\mu\nu}$ (or $\Tilde{W}_{\mu\nu}=\varepsilon_{\mu\nu\rho\sigma}W^{\rho\sigma}$) gets squared when moving from (\ref{L}) to (\ref{eff_model}).
If both \(\gamma\) and \(\Tilde{\gamma}\) are nonzero, the P-odd interaction term \(\frac{\gamma \Tilde{\gamma}}{2}  W^{\mu\nu} \tilde{W}^{\mu\nu}\) should also be added to the vector field Lagrangian.
In this paper, we will be interested in configurations supporting the electric (or magnetic) field in the center-of-mass frame, hence the latter term is identically zero for our solitons.

The theory (\ref{eff_model}) possesses a conserved current corresponding to the global $U(1)$ symmetry,
\begin{equation}
j^\nu = i (V^{*\mu\nu}V_{\mu} - V^{\mu\nu}V^*_{\mu}) \;.
\end{equation}
Besides, there is a ``topological'' current
\begin{equation}
j_{T}^\nu = i \partial_\mu W^{\mu\nu} \;,
\end{equation}
which is manifestly conserved due to the antisymmetry of $W_{\mu\nu}$.
The $U(1)$-current defines the global charge $Q$ of solitonic solutions; the classically stable (unstable) soliton provides a local minimum (maximum) of the Hamiltonian of the theory at a fixed value of $Q$ \cite{Coleman:1985ki}.
As for the current $j_{T}^\nu$, the associated charge will vanish identically on the solutions studied below.

The Einstein energy-momentum tensor of the theory (\ref{eff_model}) takes the form
\begin{multline}
T_{\mu\nu} = - V^*_{\lambda\mu}V^\lambda_\nu - V_{\lambda\mu}V^{*\lambda}_\nu \\ + g_{\mu\nu}\frac{1}{2}V^*_{\lambda\rho}V^{\lambda\rho} - \frac{1}{2}\left(\frac{\partial \Tilde{U}}{\partial V^\mu} V_\nu + \frac{\partial \Tilde{U}}{\partial V^\nu} V_\mu \right) \\ - \frac{1}{2}\left(\frac{\partial \Tilde{U}}{\partial V^{*\mu}} V^*_\nu + \frac{\partial \Tilde{U}}{\partial V^{*\nu}} V^*_\mu \right) + \Tilde{U} \;,
\end{multline}

and the energy density is
\begin{equation}
T_{00} = V^*_{0i}V_{0i} + \frac{1}{2} V^*_{ij}V_{ij} - \frac{\partial \Tilde{U}}{\partial V_0} V_0 - \frac{\partial \Tilde{U}}{\partial V^*_0} V^*_0 + \Tilde{U} \;.
\end{equation}
The latter determines the energy $E$ of the solitonic solutions.
Note that $T_{00}$ coincides with the energy density of the original theory (\ref{L}) provided that \cref{E_to_W2} is fulfilled, that is, if there are no extra sources of the electromagnetic field.

\section{Classical solutions in vector theory}
\label{sec:sol}

In this section we study nontopological vector solitons in the theory (\ref{eff_model}).
By \cref{E_to_W2}, any spatially localised solution with nonzero $W^{\mu\nu}$ can host the electric (magnetic) field that is confined to the bulk of the solution and vanishes exponentially fast outside it.
As explained above, we are interested in the solitons that admit the nonrelativistic limit.
Besides, we look for solutions that are kinematically stable against dissolution into a gas of free particles.
This is ensured by requiring that $E<MQ$.

To facilitate the study, we switch to the dimensionless units as follows,
\begin{equation}
\label{var}
V^\nu = \frac{M\Tilde{V}^\nu}{\sqrt{|\Tilde{\alpha}|}} \;, \,\,\, \kappa = \frac{\Tilde{\beta}}{|\Tilde{\alpha}|} \;, \,\,\, x_ix^i = \frac{r^2}{M^2} \;, \,\,\, t = \frac{\tau}{M} \;.
\end{equation}
The action $S$ of the theory becomes $S=|\Tilde{\alpha}|^{-1}\tilde{S}$ where $\Tilde{S}$ depends only on $\kappa$. As we will see, the physically interesting region of parameters is $|\kappa|\sim 1$.
The validity of the semiclassical approach requires $|\Tilde{\alpha}|/4\pi\ll 1$.

In units (\ref{var}), the vector field equations become
\begin{equation}
\label{vectorEOM}
\partial_\mu \Tilde{V}^{\mu\nu} + \Tilde{V}^\nu \pm (\Tilde{V}^*_\mu\ \Tilde{V}^\mu) \Tilde{V}^\nu + \kappa (\Tilde{V}_\mu\Tilde{V}^\mu) \Tilde{V}^{*\nu} = 0 \;,
\end{equation} 
where the sign of the third term coincides with the sign of $\tilde{\alpha}$.

\subsection{Condensate}
\label{ssec:cond}

Before studying solitons, it is useful to consider spatially homogeneous solutions of \cref{vectorEOM}---vector condensates.
There are two types of relevant condensates. The first one is described by the following solution,
\begin{equation}
\label{condensate}
\Tilde{V}_0 = 0, \,\,\
\Tilde{V}^1 = \sqrt{\frac{1-\mathrm{w} ^2}{\pm 1 + \kappa}} \: \e^{-i\mathrm{w} \tau} \;,  \,\,\ \Tilde{V}^i  = 0 \;, \,\ i = 2,3
\end{equation}
Here w is the angular frequency in units of the field mass, and the sign in the denominator follows the sign of $\tilde{\alpha}$.
In the non-relativistic limit, $ \mathrm{w}\to 1$, this condensate approaches the classical ground state $\Tilde{V}^\mu=0$.
At $\mathrm{w}<1$, the solution (\ref{condensate}) describes weakly-interacting vector particles in a spatially-homogeneous bound state.
The existence of such bound state is important for the non-relativistic solitons that are physically formed out of the gas of weakly-interacting particles.
Thus, we require \(\pm 1 + \kappa > 0\) for our solitons.
Another way to obtain this constraint is to study directly the non-relativistic limit of solitons with a particular ansatz; see, e.g., \cite{Loginov:2015rya}.

The second condensate that we consider has non-zero \(W^{01}\) and \(F^{01}\), and it allows us to further narrow the range of physically interesting parameters. The ansatz is qualitatively similar to the one used in \cite{Loginov:2015rya}:
\begin{equation}
\label{condensate2}
\Tilde{V}_0 = i u \e^{-i\mathrm{w} \tau} \;, \,\,\
\Tilde{V}^1 = v \e^{-i\mathrm{w} \tau} \;,  \,\,\ \Tilde{V}^i  = 0 \;, \,\ i = 2,3
\end{equation}
where \(u\) and \(v\) are real constants depending on \(\mathrm{w}\). 
Using equations of motion (\ref{vectorEOM}), we obtain that the bound-state solution of the form (\ref{condensate2}) exists at $-1<\kappa<0$.
The existence of this bound state is important for the kinematic stability of the solitons.
Indeed, the solitons with $E<MQ$ are often associated with the thin-wall limit \cite{Coleman:1985ki}.
Near this limit, the bulk of the soliton, which makes the dominant contribution to its charge and energy, is close to the respective condensate solution.
If the latter is kinematically stable, so is the soliton.
As we will see, the ansatz (\ref{condensate2}) describes the bulk of our solitons in the thin-wall regime.
Thus, we require $-1<\kappa<0$, which, together with the bound \(\pm 1 + \kappa > 0\), also implies that $\tilde{\alpha}>0$.

\subsection{Solitons}
\label{ssec:soliton}

Consider the following spherically-symmetric radial ansatz for the vector field (see \cite{Loginov:2015rya} and references therein): 
\begin{equation}
\label{ansatz}
\Tilde{V}_0 = i u(r) \e^{-i\mathrm{w} \tau} \;, \,\,\
\Tilde{V}^i = \frac{\tilde{x}^i}{r} v(r) \e^{-i\mathrm{w} \tau} \;,\,\,\,
\end{equation}
where $\Tilde{x}^i = Mx^i$, \(u(r)\), \(v(r)\) are profile functions of the vector field, and \(\mathrm{w}\) is the angular frequency in units of the field mass.
Substituting to \cref{vectorEOM}, we obtain
\begin{equation}
\label{ans_eq}
    \begin{aligned}
&u'' + \frac{2}{r} u' - \mathrm{w}(v' + \frac{2}{r} v) \\
& \qquad\qquad - u - (u^2 - v^2)u - \kappa (u^2 + v^2) u = 0 \;, \\
& \mathrm{w} u' + (1 - \mathrm{w}^2) v \\
& \qquad\qquad + (u^2 - v^2)v - \kappa (u^2 + v^2) v = 0 \;,
    \end{aligned}
\end{equation}
where prime means the derivative with respect to $r$, and the sign of the cubic term is fixed by the requirement $\tilde{\alpha}>0$.
We solve these equations numerically with the appropriate boundary conditions for $u$ and $v$.
The latter are set by the regularity at the origin ($u'\to 0$, $v\to 0$, $v'\to \mathrm{const}$, $r\to 0$) and approaching the classical ground state at infinity ($u,v\to 0$, $r\to\infty$).

With the ansatz (\ref{ansatz}), the global $U(1)$ charge and energy of the soliton are given by
\begin{align}
\label{QQ}
    & Q = \frac{8\pi}{|\Tilde{\alpha}|} \limitint_0^\infty v(\mathrm{w} v - u' )r^2 \: \diff r \;, \\ 
    & E = \frac{4\pi M}{|\Tilde{\alpha}|} \limitint_0^\infty  \bigl( (\mathrm{w} v - u')^2 \nonumber + \\ 
\label{EE}
    & \quad + \bigl( u^2 + v^2 \bigr) + \frac{1}{2}(u^2 - v^2)(3 u^2 + v^2)  \\
    & \quad + \frac{\kappa}{2}(u^2 + v^2)(3u^2 - v^2)\bigr)r^2 \: \diff r \;. \nonumber
\end{align}
Taking the derivatives with respect to $\omega\equiv M$w in these expressions and using the equations of motion, it is straightforward to show that
\begin{equation}
    \label{dEdQ}
\frac{\partial E}{\partial\omega} = \omega\frac{\partial Q}{\partial \omega} \;,
\end{equation}
which is the well-known general relation for non-topological solitons.
We use it to check the consistency of our numerical routines.

The rest of the paper is dedicated to solving numerically \cref{ans_eq} with the solitonic boundary conditions, and discussing the properties of the obtained solutions.
Our procedure is as follows.
We fix the value of $\kappa$ from the range $-1<\kappa<0$, in which one expects solutions both in the non-relativistic and the thin-wall (kinematically stable) limits.
Using the shooting method, we first find solutions close to the non-relativistic limit, \(\mathrm{w} \rightarrow 1\). 
In this limit, the field amplitudes tend to zero, and the validity of the effective theory is ensured.

\begin{figure}[t]
\includegraphics[width=0.9\linewidth, height=5cm]{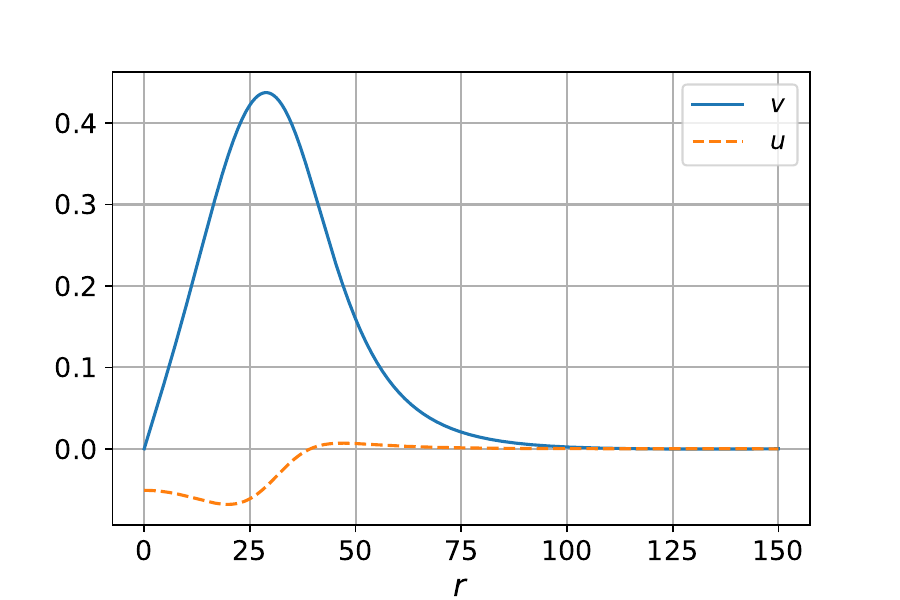}
\caption{ Vector soliton (\ref{ansatz}) at \(\kappa = - 0.9\) and \(\mathrm{w} =  0.998\).
This solution is kinematically stable, \(E < MQ\).
}
\label{kappa-0.9_w0.998}
\end{figure}

\begin{figure} [b]
\includegraphics[width=0.9\linewidth, height=5cm]{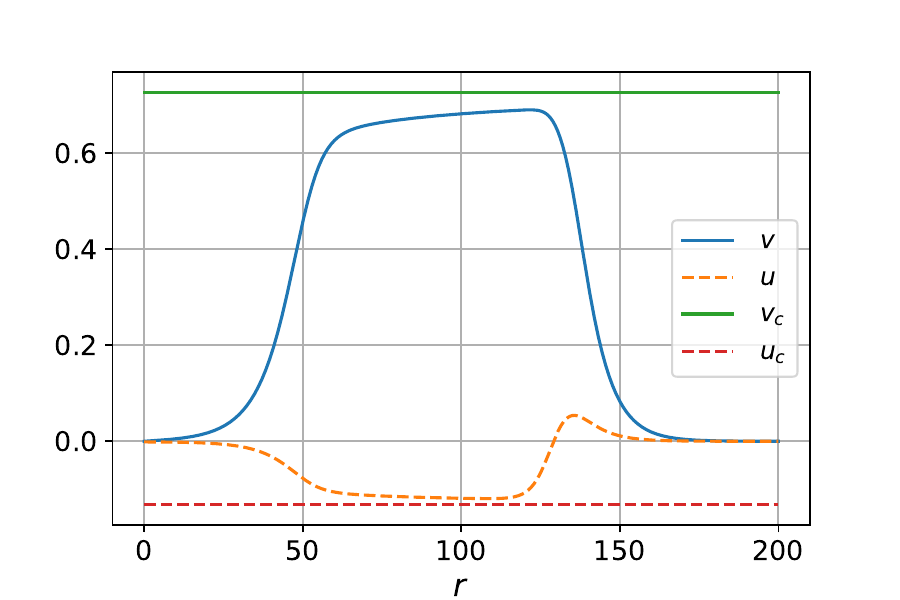}
\caption{
Kinematically stable vector soliton (\ref{ansatz}) at \(\kappa = - 0.9\) and \(\mathrm{w} =  0.99\).
Also shown is the condensate solution (\ref{condensate2}) (\(u_c\), \(v_c\)) for the same parameters.
}
\label{kappa-0.9_w0.99}
\end{figure}

Then we look for solutions with smaller w by an iterative procedure. 
We find the frequency $\mathrm{w}_s$ at which $E=MQ$; at $\mathrm{w}< \mathrm{w}_s$ the solitons are kinematically stable.
Fig.~\ref{kappa-0.9_w0.998} shows an exemplary solution with $E<MQ$ living close to \(\mathrm{w_s}\), at $\kappa=-0.9$.
Depending on the value of $\kappa$, we may enter the thin-wall regime as we further decrease w, see Fig.~\ref{kappa-0.9_w0.99} for the illustration.
The thin-wall limit corresponds to $\mathrm{w} \to \mathrm{w}_{min}>0$.

Let us discuss the limit $ \mathrm{w} \to \mathrm{w}_{min}$ in more detail.
There are two types of solutions near this limit; both of them are kinematically stable.
Solutions of the first type---the proper thin-wall solitons---are found in the range $-1<\kappa\lesssim 0.6$.
For such solitons, the region of the central depression of the fields is followed by a broad bulk where the magnitudes of $u$, $v$ approach the values of the condensate solution (\ref{condensate2}) with the same frequency. 
Outside the bulk, the fields rapidly vanish.
This structure is illustrated in Fig.~\ref{kappa-0.9_w0.99}.
The size of the soliton, its total charge and energy grow indefinitely as $\mathrm{w} \to \mathrm{w}_{min}$.

\begin{figure}[t]
\includegraphics[width=0.9\linewidth, height=5cm]{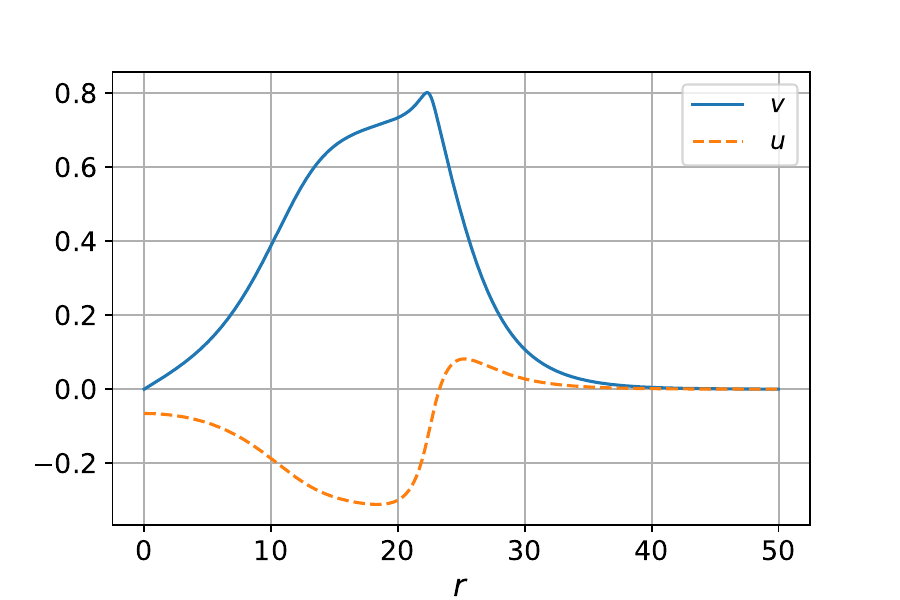}
\caption{
Kinematically stable vector soliton (\ref{ansatz}) at \(\kappa = - 0.55\) and \(\mathrm{w} = 0.96\).
We see that the cusp is approached in the transverse components of the field.
}
\label{kappa-0.55_w0.96}
\end{figure}

Solutions of the second type are found in the range $-0.6\lesssim\kappa\lesssim 0$.
They do not approach the thin-wall limit even though the condensate (\ref{condensate2}) still exists.
The obstruction to the limit lies in the structure of the differential equations (\ref{ans_eq}).
To see it, we express $v'$ in terms of \(u\), \(\chi \equiv u'\) and \(v\) (which brings \cref{ans_eq} to the standard form for the Runge-Kutta method):
\begin{equation}
\begin{split}
v' & = \frac{\mathrm{w}}{1 + (u^2 - v^2) - \kappa (u^2 + v^2) - 2v^2(1+\kappa)} \\ 
& \times \biggl(- \frac{2}{\mathrm{w}}(1 -\kappa)uv\chi + \frac{2(\chi - \mathrm{w}v)}{r}  \\ 
& -u - u (u^2 - v^2) -\kappa u (u^2 + v^2) \biggr) \;.
\end{split}
\end{equation}
Notice that the denominator can vanish at some $u,v>0$.
When this happens, the smoothness of the solution is lost.
We see indeed that $v(r)$ develops a cusp at some finite $r$ when the stiffness point is approached; see Fig.~\ref{kappa-0.55_w0.96} for illustration.
Because of the cusp we cannot reach the thin-wall regime.
In fact, the stiffness problem is common for classical vector field equations of the type studied here \cite{Mou:2022hqb,Coates:2022qia,Coates:2023swo}.
It indicates the breakdown of the effective theory (\ref{eff_model}) for solitons near the stiffness point.

\begin{figure}[t]
\includegraphics[width=0.9 \linewidth, height = 6 cm]{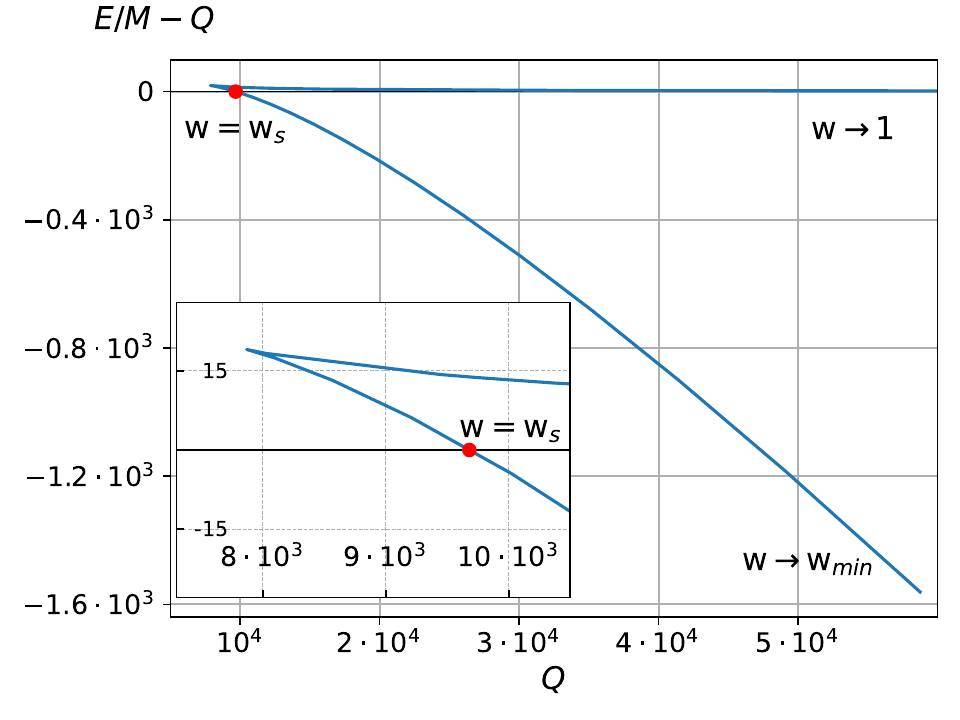}
\caption{$ E/M - Q$ as a function of $Q$ for vector solitons (\ref{ansatz}) in the theory (\ref{eff_model}) with $\tilde{\alpha} = 1$ and $\kappa = \tilde{\beta}/|\tilde{\alpha}|= -0.55$.}
\label{E - MQ}
\end{figure}

Having obtained the family of solitons parameterised by w, we can study relations between their charge $Q$ and energy $E$.
For example, Fig.~\ref{E - MQ} shows the parametric dependence of $E/M-Q$ on $Q$, at $\kappa=-0.55$.
We see the lower branch heading towards the thin-wall limit, and the upper branch connecting to the non-relativistic limit.
The branches are joined at a point where $\partial E/\partial\omega=\partial Q/\partial\omega=0$.
The qualitatively same dependencies are obtained for other allowed values of $\kappa$ (for \(\kappa \gtrsim - 0.4\), however, the soliton develops the cusp at $E/M-Q>0$, and the kinematically stable region is inaccessible).
The behavior of $E$ and $Q$ is quite similar to the one in the Friedberg-Lee-Sirlin model of Q-balls made of two scalar fields \cite{Friedberg:1976me}.

Importantly, when \(\kappa \rightarrow - 1\), the frequency $\mathrm{w}_s$, at which the solitons become kinematically stable, tends to 1.
This means that we can find solutions which are both non-relativistic and kinematically stable. The stiffness problem never occurs to them. Note that even though $\mathrm{w}$ changes very little for such solutions, the possible values of $Q$ and $E$ are not bounded from above, thanks to the thin-wall limit.
All these properties are favourable for considering such solitons as ``hosts'' of the electric (or magnetic) field.

\subsection{Applicability of classical theory}
\label{ssec:corr}
  
Let us discuss the validity of the solutions obtained above. 
We work in the regime 
$\gamma\ll \alpha\lesssim 1$.
%$1\gtrsim\alpha\gg\gamma$.
One of these inequalities is required by the semiclassical approximation: as discussed above, the classical action of the theory with the potential (\ref{U1}) scales as $|\alpha|^{-1}$, hence quantum corrections are suppressed by the powers of small coupling $\alpha$ and additional loop factor $1/(4\pi)^2$. The strong coupling scale of the theory is estimated as $\Lambda_{\alpha} \sim M\sqrt{4\pi}/\alpha^{1/4}$ \cite{Porrati:2008gv}. It should be larger than the interacting energy of particles in the soliton, which is satisfied automatically for our solutions at small coupling constants. Note that, even though the magnitude of the field in the soliton may exceed the cutoff scale, \(V^{\mu} \sim M/\sqrt{\alpha} > \Lambda_{\alpha}\), this does not mean the breakdown of the semiclassical approximation, which follows from the loop expansion controlled by the powers of $\alpha$.\footnote{A similar procedure is used to validate the semiclassical analysis of the Higgs mechanism for parametrically large vacuum expectation values \cite{Coleman:1973jx}.} The second inequality is motivated by considering $V_\mu$-particles as a dark matter candidate. For example, the direct constraint on the magnetic dipole moment of electromagnetically interacting vector particles with masses of order 100 MeV is $\gamma< 10^{-8}$ \cite{Chu:2023zbo}.

The charge (\ref{QQ}) and energy (\ref{EE}) of our solutions satisfy $Q\gg 1$, $E/M\gg 1$, meaning that the solitons are composed of many quanta of the field $V^\mu$.
Furthermore, the size of the soliton is much larger than its Compton wavelength $\sim E^{-1}$.
Finally, from Fig.~\ref{E - MQ} we see that $|E-MQ|\ll MQ$, meaning that the binding energy of $V^\mu$-particles inside the solitons is small compared to their rest energy, and the particles are non-relativistic.

Note, finally, that the solitons with the cusp, such as the one presented in Fig.~\ref{kappa-0.55_w0.96}, are beyond the applicability limit of the effective theory, since they have large quantum corrections \cite{Mou:2022hqb}.

\section{electromagnetic hedgehogs}
\label{sec:hedgehog}

\begin{figure}[b]
\includegraphics[width=0.92 \linewidth, height=6 cm]{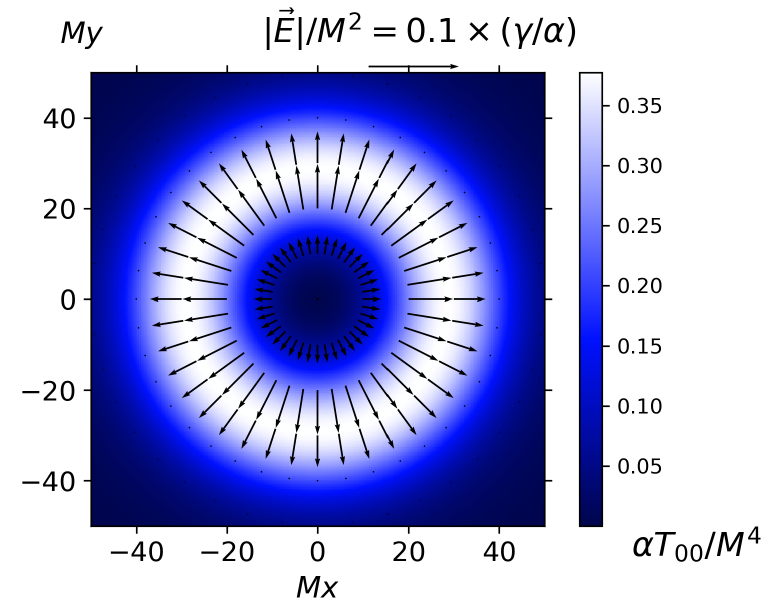}
\caption{
The equatorial cross-section of the electric hedgehog---radial spherically-symmetric nontopological vector solition---in the theory (\ref{L}).
We take \(\kappa = \beta / \alpha = -0.9\), \( \mathrm{w} = 0.998\), and \(\gamma \ll \alpha\).
The arrows indicate the value \(\vert \vec{E}\vert/M^2\) and the direction of the electric field.}
\label{hedgehog}
\end{figure}

According to \cref{E_to_W2}, the vector solitons can confine the electromagnetic field. 
As discussed above, we choose $\gamma\ll\alpha\lesssim 1$.
At $-1<\kappa< 0$, we have found solitons which are non-relativistic, kinematically stable, and have a wide range of physical parameters $Q$ and $E$.
These spherically-symmetric solutions support radial fields---``hedgehogs'', see Fig.~\ref{hedgehog} for illustration.
For the P-even interaction between $F_{\mu\nu}$ and $W^{\mu\nu}$, it is the radial electric field trapped in the bulk of the soliton, while for the dual theory (\(\gamma = 0 \), \( \Tilde{\gamma} \neq 0\)) it is the radial magnetic field.
Note that these nontopological solitons are not magnetic monopoles, since the magnetic field decreases exponentially fast at large distances.
This localisation of the field is the feature of the theory (\ref{L}).

Fig.~\ref{hedgehog} shows the solution with \(\kappa = \beta / \alpha = -0.9\), \( \mathrm{w} = 0.998\).
The solution confines the electric field in its central ring-like region.
The maximum of the field strength in SI units ($\mathrm{V}/\mathrm{m}$) is
\begin{equation}
\vert \vec{E}\vert  \sim \frac{\gamma}{\alpha} \times 10^{5} \frac{M^2}{\mathrm{eV^2}} \frac{\mathrm{V}}{\mathrm{m}} \;.
\end{equation}

Since $\mathrm{w} \approx 1$, the solution can be described in the non-relativistic language where it has a clear physical interpretation.
In the non-relativistic regime, one can re-write the equation of motion for $V^i$ in the form of the Gross-Pitaevsky equation with additional terms describing magnetic dipole and electric quadrupole interactions, and use the transversality condition $\partial_\mu V^\mu = 0$ to exclude $V_0$ \cite{Pauli:1941zz}.
Note that the magnetic dipole moment vanishes on the ansatz (\ref{ansatz}) due to antisymmetry of $\epsilon_{ijk}$. This means that the leading electromagnetic interaction is between the electric field and the electric quadrupole moment of the vector field. It is this interaction that provides the soliton with the structure of the electric hedgehog shown in  Fig.~\ref{hedgehog}.

%{\bf Our configuration has clear interpretation in the non-relativistic approximation. Indeed, the choice of $\mathrm{w} \to 1$ allows to re-wright E.O.M. for the model (\ref{L}) in a form of Gross-Pitaevsky equations
%with additional terms which describe magnetic dipole and electric quadrupole interactions.
%In this case
%$V_0$ could be excluded through the transversality condition $\partial_\mu V^\mu = 0$, which becomes valid in the non-relativistic limit \cite{Pauli:1941zz}. 
%Furthermore, on the substitution  (\ref{ansatz}) for spatial components of vector fields magnetic dipole moment is equal to zero due to antisymmetricity of $\epsilon_{ijk}$. Thus, the leading electromagnetic term is that describes interaction of electric field with the quadrupole moment of vector field. This is the interaction which provides the electric hedgehog structure of solitons (see Fig.\ref{hedgehog}).
%}
\section{Conclusion}
\label{sec:concl}

In this work, we studied solitons in a theory of complex vector field $V^\mu$ coupled to the electromagnetic field as in \cref{L}.
We obtained kinematically stable solutions close to the nonrelativistic limit that can confine radial electric or magnetic field in their interior.
Importantly, we worked in the limit in which the coupling of $V^\mu$ to the electromagnetic field is much smaller than other couplings in the theory. 
In cosmology, this is motivated by the fact that $V^\mu$ can be a dark matter candidate.
On the other hand, the theory (\ref{L}) can also be used as a mean field description of a Bose-Einstein condensate of atoms or molecules with the large dipole moment $\sim\gamma/M$ \cite{Lahaye:2009qqr}.
The classical solutions found here are readily adapted to larger values of $\gamma$.

Let us outline possible directions for future work.
One interesting generalisation of the present study would be to consider the coupling of the global vector field to a non-abelian gauge field \cite{Glashow:1959wxa}.
The gauge field lies in the adjoint representation, and one can consider the dipole interaction for the neutral vector field $V$ in the fundamental representation.
The interaction term
\[
i \Tr(V^\dagger_\mu F_{\mu\nu} V_{\nu}) + h.c.
\]
is invariant under $U_g(1)$ and the gauge group. 

Another natural generalisation is to use gravity as a binding force of the vector soliton, rather than the field self-interaction.
This would allow one to disentangle the parameters of the soliton from its coupling to the electromagnetic field.
Next, one can add a Higgs-like field in the theory (\ref{eff_model}) to obtain configurations with lower values of \(\mathrm{w}\).
Solitons in a resulting theory---``Proca-Higgs balls'' and their gravitating counterparts---are a subject of recent research \cite{Herdeiro:2023lze,Brito:2024biy}.
Finally, one can study small perturbations of the solutions presented here. 
In particular, it is important to analyse linear classical stability of the vector solitons and of the electromagnetic field trapped by the soliton. 

\section{Acknowledgments}\label{acknowledgements}
The authors are grateful to Eduard Kim, Yakov Shnir, Sergey Troitsky for useful discussions and helpful comments on the paper. 
Numerical studies of this work were supported by the grant RSF 22-12-00215.
Research at Perimeter Institute is supported in part by the Government of Canada
through the Department of Innovation, Science and Economic Development Canada and by
the Province of Ontario through the Ministry of Colleges and Universities.

\bibliography{biblio}% Produces the bibliography via BibTeX.

\end{document}